\documentclass[twocolumn, superscriptaddress, prb, nofootinbib]{revtex4}
\usepackage{graphicx}
\usepackage{amsmath}
\usepackage{multirow}
\usepackage{textcomp}
\usepackage{float}
\usepackage{hyperref}
\usepackage[symbol]{footmisc}
\usepackage{perpage}
\begin{document}
\title{
Influence of inversion on Mg mobility and electrochemistry in spinels
}

\author{Gopalakrishnan Sai Gautam} \email{gautam91@mit.edu}
\affiliation{
Department of Materials Science and Engineering, Massachusetts
Institute of Technology, Cambridge, MA 02139, USA}
\affiliation{
Materials Sciences Division, Lawrence Berkeley National Laboratory,
Berkeley, CA 94720, USA}
\thanks{G. S. Gautam and P. Canepa contributed equally to the work}

\author{Pieremanuele Canepa}
\thanks{G. S. Gautam and P. Canepa contributed equally to the work}
\affiliation{
Materials Sciences Division, Lawrence Berkeley National Laboratory,
Berkeley, CA 94720, USA}

\author{Alexander Urban}
\affiliation{
Materials Sciences Division, Lawrence Berkeley National Laboratory,
Berkeley, CA 94720, USA}

\author{Shou-Hang Bo}
\affiliation{
Materials Sciences Division, Lawrence Berkeley National Laboratory,
Berkeley, CA 94720, USA}

\author{Gerbrand Ceder} \email{gceder@berkeley.edu, gceder@lbl.gov}
\affiliation{
Department of Materials Sciences and Engineering, University of California Berkeley,
CA 94720, USA}
\affiliation{
Materials Sciences Division, Lawrence Berkeley National Laboratory,
Berkeley, CA 94720, USA}



\begin{abstract}
Magnesium oxide and sulfide spinels have recently attracted interest as cathode and electrolyte materials for energy-dense Mg batteries, but their observed electrochemical performance depends strongly on synthesis conditions.
Using first principles calculations and percolation theory, we explore the extent to which spinel inversion influences Mg$^{2+}$ ionic mobility in MgMn$_2$O$_4$ as a prototypical cathode, and MgIn$_2$S$_4$ as a potential solid electrolyte.
We find that spinel inversion and the resulting changes of the local cation ordering give rise to both increased and decreased Mg$^{2+}$ migration barriers, along specific migration pathways, in the oxide as well as the sulfide. 

To quantify the impact of spinel inversion on macroscopic Mg$^{2+}$ transport, we determine the percolation thresholds in both MgMn$_2$O$_4$ and MgIn$_2$S$_4$. Furthermore, we analyze the impact of inversion on the electrochemical properties of the MgMn$_2$O$_4$ cathode via changes in the phase behavior, average Mg insertion voltages and extractable capacities, at varying degrees of inversion.
Our results confirm that inversion is a major performance limiting factor of Mg spinels and that synthesis techniques or compositions that stabilize the well-ordered spinel structure are crucial for the success of Mg spinels in multivalent batteries.
\end{abstract}

\maketitle
\section{Introduction}
\label{sec:intro}
Multivalent (MV) batteries, such as those based on Mg$^{2+}$,{\cite{Aurbach2000,Canepa2017}} can potentially achieve high volumetric energy density via facile non-dendritic stripping/deposition on an energy-dense metal anode.{\cite{Yoo2013a,Besenhard2002,Muldoon2014}} However, the development of viable MV technology is hindered by poor Mg diffusivity in oxide cathodes as well as poor Coulombic efficiencies in liquid electrolytes.{\cite{Canepa2017,Muldoon2014,Canepa2015a,Canepa2015}}

One pathway to improve Mg migration in solids is to utilize host structures where Mg occupies an unfavorable coordination environment.{\cite{Rong2015,Brown1988,Gautam2015}} Spinels with composition AM$_2$X$_4$ (A~=~Mg, M~=~metal cations, X~=~O or S) are appealing structures in this regard because of their tetrahedrally-coordinated Mg sites, rather than the preferred octahedral coordination of Mg. Theoretical calculations indeed predict reasonable Mg$^{2+}$ migration barriers ($\sim 550 - 750$~meV) in both oxide and sulfide spinels.{\cite{Liu2015,LiuJainQuEtAl2016}}  Note that oxide spinels have long been used as cathodes and anodes in commercial Li-ion batteries.{\cite{Whittingham2004,Thackeray1984,Wagemaker2005,Reed2001,Anton2000_SpinelMnO2,Ferg1994,Andre2015}}

Spinel-Mn$_2$O$_4$ is a particularly promising, energy-dense, MV cathode, as it is one of the few oxides{\cite{Gershinsky2013,SaiGautam2015,SaiGautam2016,Sa2016,Tepavcevic2015,Novak1994}} to have shown electrochemically reversible Mg$^{2+}$ intercalation.{\cite{Kim2015a,FengChenQiaoEtAl2015}}  However, the cyclable Mg content, i.e., the observed capacity, seems to depend strongly on the synthesis conditions.{\cite{Kim2015a,FengChenQiaoEtAl2015,Kim2015b}} Several studies on the MgMn$_2$O$_4$ structure{\cite{Irani1962,Malavasi2002,Rosenberg1964,Manaila1965,Radhakrishnan1976}} have indicated that the spinel is prone to \emph{inversion}, i.e., Mg/Mn antisite disorder (see Section~{\ref{sec:Structure}}), where the \emph{degree of inversion} can range from 20\%{\cite{Malavasi2002}} to 60\%.{\cite{Irani1962}} It has further been argued that the propensity of Mn$^{3+}$ to disproportionate into Mn$^{2+}$ and Mn$^{4+}$ promotes spinel inversion and phase transformations.{\cite{Reed2001,Reed2004}}  Since inversion directly affects the local cation arrangement, it may significantly impact the Mg$^{2+}$ ionic mobility.{\cite{Lee2014a,Urban2014}}  For the rational design of improved Mg battery cathodes it is, therefore, crucial to understand how inversion in oxide spinels affects Mg$^{2+}$ migration.

Inversion is not a phenomenon unique to oxides, and other chalcogenide spinels such as sulfides, which are also important cathode materials in MV technology,{\cite{LiuJainQuEtAl2016}} are also known to exhibit inversion.{\cite{Seminovski2012,Canepa2017a}} A recent combined theoretical and experimental study has identified ternary sulfide and selenide spinels as promising Mg-ion conductors with potential applications as solid electrolytes in MV batteries.{\cite{Canepa2017a}}  Solid electrolytes combine the advantage of improved safety with a high Mg transference number.  Three promising compounds were reported, namely, MgSc$_2$Se$_4$, MgSc$_2$S$_4$, and MgIn$_2$S$_4$.{\cite{Canepa2017a}} MgIn$_2$S$_4$ spinel had previously been reported,{\cite{Hahn1950,Wakaki1980}} and the available literature as well as our own synthesis attempts (Figure~S1 in Supporting Information, SI\footnote{$^\#$ Electronic Supporting Information available free of charge online at \url{http://dx.doi.org/10.1021/acs.chemmater.7b02820}}$^{\#}$) indicate that the compound is prone to inversion, where the degree of inversion can be as high as $\sim$~85\% (Table~S2 in SI).

In the present work, motivated by the importance of the spinel structure for MV battery technology, we explore the influence of spinel inversion on Mg mobility in ternary oxides and sulfides, using MgMn$_2$O$_4$ and MgIn$_2$S$_4$ as the prototype for each class of spinels.  We consider all possible local cation environments that arise due to inversion and compute the activation barriers for Mg migration in each scenario using first-principles calculations. 
The high requirement for the ionic conductivity in solid electrolytes typically demands migration barriers to be $<~500$~meV, as observed in solid Li-conductors,{\cite{BachmanMuyGrimaudEtAl2016}} while cathodes can operate under lower ionic mobilities (barriers $\sim~750$~meV, see Section~{\ref{sec:results_MnO2}}){\cite{Canepa2017}} as the required length is less than for a conductor. Hence, we limit accessible Mg$^{2+}$ migration paths to those with a barrier less than 500~meV and 750~meV for operation as a solid electrolyte and cathode, respectively. We will use MgMn$_2$O$_4$ as the prototype cathode for which we restrict barriers to 750~meV and MgIn$_2$S$_4$ as an example of an electrolyte (barriers $<~500$~meV).

Our results indicate that inversion, in both solid electrolytes and cathodes, can simultaneously cause a decrease in activation barriers across certain migration trajectories while increasing the barriers across others, leading to a complex interplay of opening and closing of specific Mg migration pathways.  To quantify the impact of these variations in the microscopic activation barriers on macroscopic Mg diffusion, we estimate the critical Mg concentrations (percolation thresholds) required to facilitate Mg$^{2+}$ diffusion through the structure at different degrees of inversion.  Note that Mg extraction from the cathode material creates Mg-vacancies that can affect the percolation properties. For example, vacancies can cause migration pathways that are inactive in the fully discharged composition to become accessible. Hence, for a cathode, we examine the variation of the percolation threshold with vacancy content in the spinel lattice. In electrolytes, the Mg concentration does not significantly vary and we do not consider the effect of Mg-vacancies in MgIn$_2$S$_4$. Our estimates indicate that stoichiometric MgMn$_2$O$_4$ and MgIn$_2$S$_4$ spinels remain percolating up to $\sim$~55--59\% and 44\% inversion, respectively.  Finally, we discuss the impact of spinel inversion on Mg-electrochemistry in the Mn$_2$O$_4$ cathode by evaluating the 0~K phase diagram, average voltages and the accessible Mg capacity at various degrees of inversion.

While previous studies have analyzed the impact of inversion on structural, thermal, electronic, and magnetic properties,{\cite{Walsh2007,Das2016,Schwarz2015,Santos-Carballal2015,Malavasi2002}} the effect on Mg mobility in spinels has not yet been investigated. Understanding the influence of inversion on ion mobility will provide guidelines to tune the synthesis and electrochemical conditions of both cathodes and solid electrolytes, not only in MV systems but also in existing Li-ion architectures.{\cite{Xu2010a}}  Finally, our results emphasize the importance of the topology of cation sites in setting the migration behavior within a general anion framework.{\cite{Gautam2016}}


\section{Structure}
\label{sec:Structure}
A spinel configuration is a specific ordering of cation sites (A and M in AM$_2$X$_4$) in a face-centered cubic (FCC) packing of anion sites (X), as shown in Figure~{\ref{fig:1}}. In a ``normal" spinel, half of the octahedral ($oct$) sites, i.e., $16d$, are occupied by M atoms (Mn/In, blue octahedra in Figure~{\ref{fig:1}), while 1/8 of the tetrahedral ($tet$) sites ($8a$) are occupied by A (Mg, orange tetrahedra) cations.

\begin{figure*}[t]
\includegraphics[width=\textwidth]{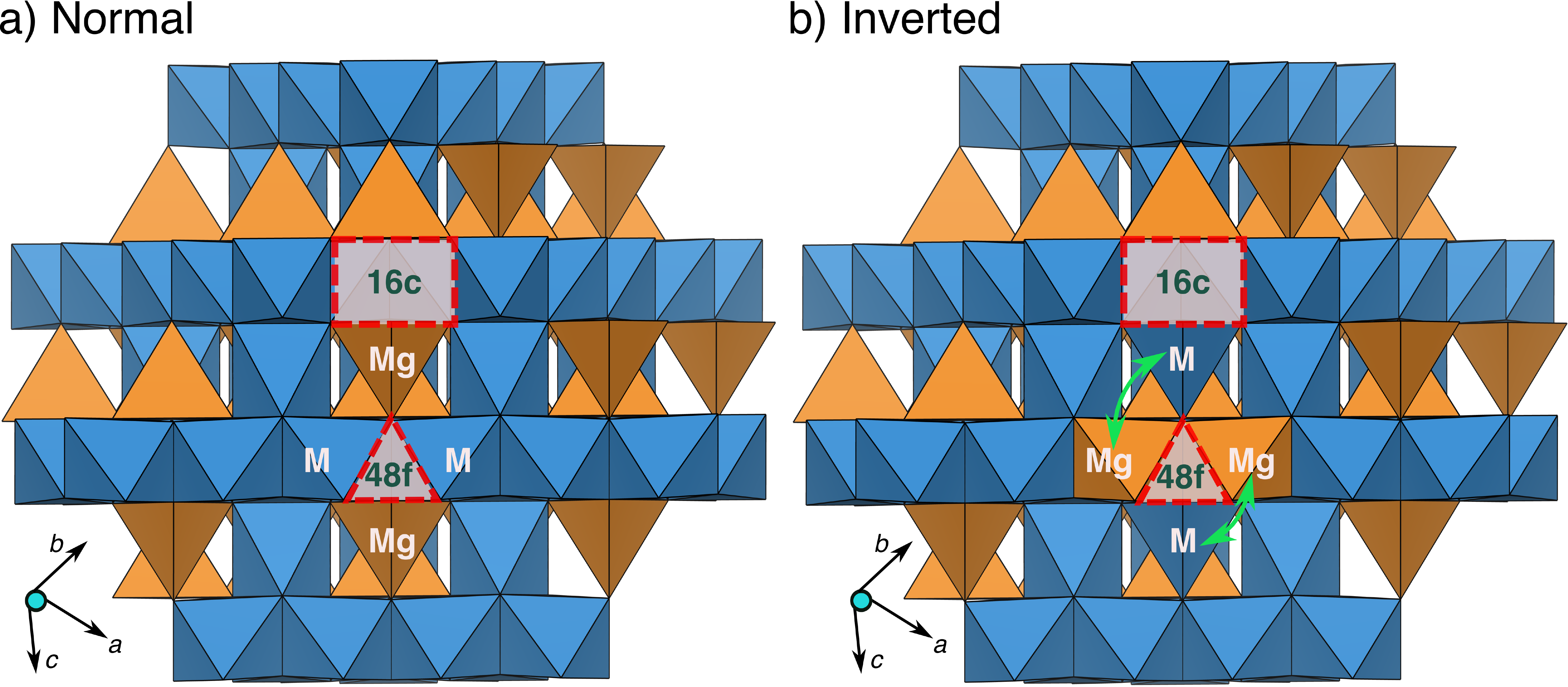}
\caption{Schematic of a normal (a) and an inverted (b) spinel MgM$_2$X$_4$ (M~=~Mn, In and X~=~O, S). The blue and orange polyhedra correspond to the M ($16d, oct$) and Mg ($8a, tet$). The dashed rectangle indicates the vacant $16c, oct$ site and the dashed triangle  the vacant $48f~tet$ site. In (b),  green arrows display the exchange of Mg and M sites, leading to inversion in the spinel.
\label{fig:1}
}
\end{figure*}

Polyhedra in the spinel structure share faces, edges and corners, as summarized in Table~{\ref{tb:notations}}. For example, the $8a$ sites that are occupied by A are face-sharing with vacant (Vac) $16c~oct$ sites (dashed red square in Figure~{\ref{fig:1}}a), edge-sharing with vacant $48f~tet$ (dashed red triangle) and corner-sharing with vacant $tet~(48f, 8b)$ and M-containing $16d$ $oct$ sites.{\cite{Sickafus1999}}  Face-sharing polyhedra have the lowest cation--cation distance, leading to the highest level of electrostatic repulsion, followed by edge-sharing and subsequently corner-sharing polyhedra.{\cite{Pauling1960}} Indeed, the $16c, 48f$ and $8b$ sites are vacant in spinel lattices ($8b$ not shown in Figure~{\ref{fig:1}}) since they face-share with occupied $8a$ or $16d$ sites.

\begin{table*}[th]
\caption{\label{tb:notations} Notations used in the AM$_2$X$_4$ structure of Figure~{\ref{fig:1}}.~Vac indicates vacancy. No.\ sites is normalized against the conventional (cubic) cell of a normal spinel with 32 anions.}
%
\begin{tabular*}{\textwidth}{@{\extracolsep{\fill}}lcccccc@{}}
\hline\hline
Site & Coordination & Ion in normal spinel & \multicolumn{3}{c}{Sharing neighbors} & No.\ sites \\
\hline
& & & Face & Edge & Corner & \\
\hline
$8a$ & $tet$ & A (Mg$^{2+}$) & $16c$ & $48f$ & $48f, 16d, 8b$ & 8 \\

$16d$ & $oct$ & M (Mn$^{3+,4+}$/In$^{3+}$) & $8b, 48f$ & $16c, 16d$ & $8a, 48f$ & 16 \\

$16c$ & $oct$ & Vac & $8a, 48f$ & $16d, 16c$ & $8b, 48f$ & 16 \\

$48f$ & $tet$ & Vac & $16d, 16c$ & $8a, 8b, 48f$ & $8a, 8b, 16c, 16d$ & 48 \\

$8b$ & $tet$ & Vac & $16d$ & $48f$ & $48f, 16c, 8a$ & 8 \\
\hline \hline
\end{tabular*}
\end{table*}

Inversion in a spinel structure refers to the collection of anti-site defects in the $8a$ (A) and $16d$ (M) sub-lattices, as shown in Figure~{\ref{fig:1}}b. The degree of inversion, $i$, is  defined as the fraction of $8a$ sites occupied by M cations, with a value of 0 (or 0\%) and 1 (100\%) indicating a normal and a fully inverted spinel, respectively.  Thus, cations A and M  are exchanged in inverted spinels (green arrows in Figure~{\ref{fig:1}}b), leading to a stoichiometry of A$_{1-i}$M$_i$[A$_{i/2}$M$_{1-(i/2)}$]$_2$X$_4$, compared to  AM$_2$X$_4$  in normal spinels.

\subsection{Possible Mg-hops}
\label{sec:Mg_hops}
Figure~{\ref{fig:2}} and Table~{\ref{tb:hops}} summarize the possible local cation arrangements in a spinel structure that can originate from inversion. The orange, blue, and green polyhedra in Figure~{\ref{fig:2}} correspond to Mg, M, and mixed (Mg/M) occupation, respectively, with the arrows in each panel indicating the Mg migration trajectory. The dashed rectangles and triangles signify vacancies. Grey polyhedra correspond to $8a$ sites that are either cation occupied or vacant. While Figure~{\ref{fig:2}}a indicates the migration trajectory in a normal spinel, panels b, c, d, and e depict the possible Mg-hops that can occur in an inverted spinel. The sub-panels in Figure~{\ref{fig:2}}b correspond to slices along perpendicular directions, i.e., the $8a$ sites in the left sub-panel of Figure~{\ref{fig:2}}b are perpendicular to the plane of the paper in the right sub-panel.

In a normal spinel, the rate for Mg diffusion is determined by the hop between adjacent $8a~tet$ sites face-sharing with a $16c$ octahedron, as shown in Figure~{\ref{fig:2}}a. Hence, the migration topology is $tet-oct-tet$, and referred to as ``Hop 1" in our work.   The intermediate $16c$ site in Hop 1 shares edges with six $16d~oct$ sites (``ring'' sites) that are occupied by M cations (2 out of 6 ring sites are shown in Figure~{\ref{fig:2}}a). It was recently  proposed{\cite{Liu2015,Rong2015,LiuJainQuEtAl2016}} that the migration barrier in normal spinels, both oxides and sulfides, is predominantly set by the size of the shared triangular face (not shown in Figure~{\ref{fig:2}}a) between the $8a~tet$ and $16c~oct$ sites.

\begin{figure*}[t!]
\includegraphics[width=\textwidth]{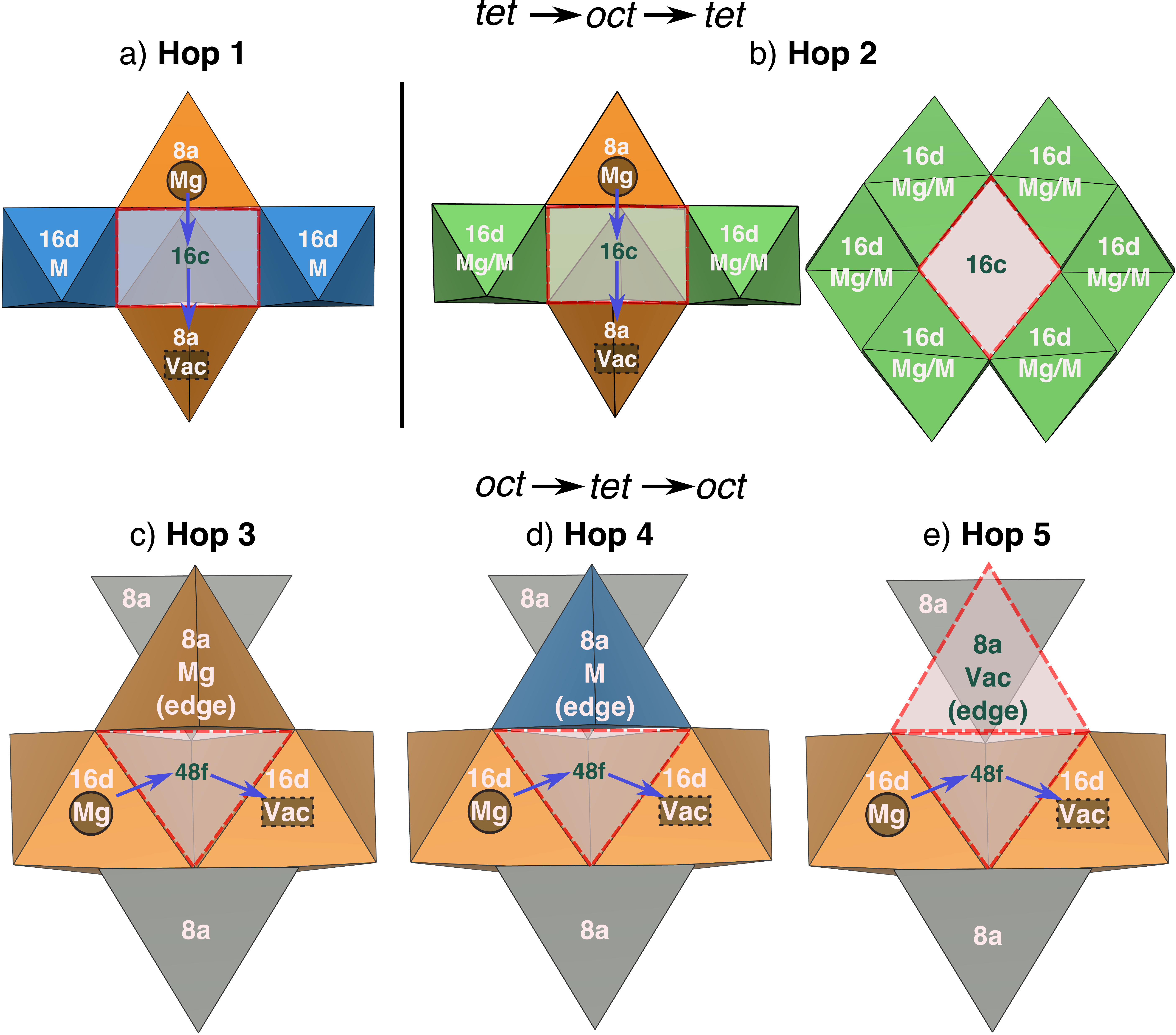}
\caption{
Local cation environments and various Mg hops considered in an inverted spinel structure. In all migration scenarios a Mg atom migrates from an occupied site (indicated by solid black circles) to an adjacent vacant site (dashed black rectangles), along the trajectory indicated by the arrows. Hops 1 (a) and 2 (b) occur with a $tet\rightarrow oct\rightarrow tet$ topology, while hops 3 (c), 4 (d), and 5 (e) occur along an $oct\rightarrow tet\rightarrow oct$ pathway. Blue and orange polyhedra correspond to Mg and M (M~=~Mn, In), while  green polyhedra indicate mixed M/Mg occupancy. In the case of Hops 3, 4, and 5 the $8a$ sites corner-sharing with the intermediate $48f$ site are shown as grey polyhedra. The notation ``edge" in panels (c), (d) and (e) corresponds to the $8a$ site that edge-shares with the $48f$. Vac indicates vacancy.
\label{fig:2}
}
\end{figure*}

Along the $tet-oct-tet$ migration pathway in inverted spinels (referred to as ``Hop 2'') the $16d$ ring sites can be occupied by both M and Mg cations, as indicated by the six green polyhedra in the right sub-panel of Figure~{\ref{fig:2}}b. To evaluate Mg$^{2+}$ migration along Hop 2, we considered multiple configurations from 1 ring site occupied by Mg to all 6 ring sites being occupied by Mg. Since each ring site occupancy (e.g., 2/6 or 3/6 Mg) corresponds to a large number of possible cation decorations on the ring sites, we used the decoration that had the lowest electrostatic energy, as obtained by minimizing the Ewald energy of the unit cell{\cite{Ewald1921}} using classical charges in the spinel framework. The specific cation arrangements used to evaluate the Mg migration barriers along Hop 2 are displayed in Figure~S13.

As inversion leads to Mg$^{2+}$ occupancy of $16d$ sites, Mg-hopping across $16d$ sites must also be considered. A $16d-16d$ hop can occur through two possible tetrahedral intermediate sites, the $8b$ and $48f$. The $8b$ sites typically share all their triangular faces with occupied $16d$ sites and are therefore not open to Mg$^{2+}$ migration due to high electrostatic repulsion, as shown by previous studies.{\cite{Gautam2016,Rong2015,Urban2014}} However, the $48f$ sites share 2 triangular faces with vacant $16c$ sites, enabling them to act as viable intermediate sites for Mg$^{2+}$ hopping. As such, we only consider the $16d-16d$ hop via the $48f$ as  intermediate site, leading to a $16d-48f-16d$ topology (Figures~{\ref{fig:2}}c, d, and e). The $48f$ shares one of its edges with an $8a~tet$ site (Table~{\ref{tb:notations}}), where the ``edge-$8a$'' can be occupied by Mg$^{2+}$ (``Hop 3", Figure~{\ref{fig:2}}c), M$^{3+/4+}$ (``Hop 4", Figure~{\ref{fig:2}}d) or a vacancy (``Hop 5", Figure~{\ref{fig:2}}c). Additionally, across Hops 3, 4, and 5, we consider two scenarios where the $8a$ sites that share a corner with the $48f$ (``corner-$8a$'', grey polyhedra in Figure~{\ref{fig:2}}) are either occupied by cations or left vacant.

\begin{table*}[th]
\caption{\label{tb:hops} Summary of all hops considered for evaluating Mg$^{2+}$ mobility in inverted spinels, where M~=~Mn, In and Vac~=~Vacancy. The neighbor column indicates the site that edge-shares with the intermediate site in the corresponding hop. The last column signifies the (maximum) number of configurations, along each migration trajectory, for which migration barriers have been calculated in this work. For example along Hop 3, the corner-$8a$ sites being cation-occupied and vacant are the two configurations considered.}
%
\begin{tabular*}{\textwidth}{@{\extracolsep{\fill}}llcc@{}}
\hline\hline
Hop & Topology & Intermediate site neighbor(s) & \# configurations \\
\hline
1 & $8a-16c-8a$ ($tet-oct-tet$) & $16d$ ($oct$, M) & 1\\
2 & $8a-16c-8a$ ($tet-oct-tet$) & $16d$ ($oct$, Mg/M) & 6\\
3 & $16d-48f-16d$ ($oct-tet-oct$) & $8a$ ($tet$, Mg) & 2\\
4 & $16d-48f-16d$ ($oct-tet-oct$) & $8a$ ($tet$, M) & 2\\
5 & $16d-48f-16d$ ($oct-tet-oct$) & $8a$ ($tet$, Vac) & 2\\
\hline \hline
\end{tabular*}
\end{table*}

\subsection{Percolation theory}
\label{sec:percolation}
While activation barriers for the various cation arrangements in Figure~{\ref{fig:2}} determine the active Mg$^{2+}$ migration hops (or channels) on the atomic scale, the macroscopic diffusion of Mg$^{2+}$, which is essential for (dis)charge of cathodes or ionic conduction in solid electrolytes, depends on the existence of a percolating network of active migration channels. As the $8a-16c-8a$ channels form a percolating network throughout the spinel structure, stoichiometric normal spinels with Mg in $8a$ enable macroscopic diffusion of Mg$^{2+}$ as long as the $8a-16c-8a$ hop is open, i.e., the migration barrier for Hop 1 is below a threshold value. However, inversion leads to mixing of cation occupancies in both the $8a$ and $16d$ sites, potentially causing some $8a-16c-8a$ channels to close (due to higher Mg$^{2+}$ migration barriers along Hop 2) while opening new channels typically closed in a normal spinel (e.g., Hops 3, 4, or 5). Hence, in addition to identifying facile microscopic hops, it is important to consider whether a percolating network of low-barrier migration channels exists. Analogous studies have been done on Li$^+$ percolation in rocksalt lattices.{\cite{Urban2014}}

In percolation theory, the site percolation problem{\cite{Stauffer1994,Hunt2006,Isichenko1992,Essam1980}} identifies the critical concentration, x~=~x$_{crit}$, at which an infinite network of contiguous connected sites exists in an infinite lattice of randomly occupied sites. In terms of ionic diffusion, x$_{crit}$ sets the ``percolation threshold", above which percolating channels exist in a given structure and macroscopic ion diffusion is feasible. While percolation thresholds are accessible analytically for 2D lattices,{\cite{Isichenko1992}} Monte-Carlo (MC) simulations need to be used to estimate x$_{crit}$ in 3D structures.{\cite{Marck1998,Lorenz1998,Urban2014}}

The existence of a percolating diffusion network in a structure at a certain x~($>$~x$_{crit}$) does not imply that all ions in the structure can be (reversibly) extracted. Mg sites that are not part of a percolating network will form isolated clusters throughout the structure so that the amount of extractable ions is lower than the total concentration, i.e., x$_{ext} <$~x. The quantity x$_{ext}$ can be assumed to correspond to the capacity of a cathode material. Numerically, x$_{ext}$ is also estimated from MC simulations.{\cite{Urban2014}}

In summary, the two central quantities obtained from percolation MC simulations are the Mg concentration beyond which macroscopic diffusion is feasible (x$_{crit}$) and the fraction of extractable Mg ions in a percolating structure (x$_{ext}$).   In order to  study Mg diffusion in spinels, we modified the nearest neighbor model (normally considered in site percolation estimations) to include occupancies up to the 3$^{rd}$ nearest neighbor (i.e., corner-sharing sites in Table~{\ref{tb:notations}}). Two Mg sites in a given spinel arrangement are considered connected only if the migration channel linking them is open (i.e., the migration barrier is below an upper-limit). Thus, a percolating network of Mg sites is formed solely via open migration channels. Whether a channel is considered open will depend on the migration barrier for Mg hopping through it. 


\section{Methods}
\label{sec:methods}
The computational approaches to predict properties relevant to cathode materials have recently been reviewed by Urban \emph{et al.}{\cite{Urban2016}} Also, the ability of Density Functional Theory (DFT){\cite{Hohenberg1964,Kohn1965}} methods to predict materials with novel properties has been amply demonstrated.{\cite{Jain2016computational}} As a result, all calculations in this work are done with DFT as implemented in the the Vienna Ab Initio Simulation Package,{\cite{Kresse1993,Kresse1996}} and employing the Projector Augmented Wave theory. \cite{Kresse1999} An energy cut-off of 520~eV is used for describing the wave functions, which are sampled on a well-converged \emph{k}-point (4$\times$4$\times$4) mesh.  The electronic exchange-correlation is described by the semi-local Perdew-Burke-Ernzerhof (PBE)\cite{Perdew1996}  functional of the Generalized Gradient Approximation (GGA). 
Calculations on Mg$_{\rm x}$Mn$_2$O$_4$ are always initialized with an ideal cubic structure while allowing for potential tetragonal distortions during the geometry relaxation as the spinel can be either cubic (x$_{\rm Mg} \sim 0$) or tetragonal (x$_{\rm Mg} \sim 1$) based on the concentration of Jahn-Teller active Mn$^{3+}$ ions. The computed $c/a$ ratio for the tetragonal-MgMn$_2$O$_4$ structure is in excellent agreement with experimental reports{\cite{Irani1962,Sanjana1960}} (see Section S12). For voltage and 0~K phase diagram calculations of Mg$_{\rm x}$Mn$_2$O$_4$, the PBESol exchange-correlation functional{\cite{Perdew2008}} is used to improve the description of the energetics,{\cite{KitchaevPengLiuEtAl2016}} while a Hubbard \emph{U} correction of 3.9~eV is added to remove spurious self-interaction of the Mn \emph{d}-electrons.\cite{Anisimov1991,Zhou2004,Jain2011}

The activation barrier calculations are performed with the Nudged Elastic Band (NEB) method.{\cite{Henkelman2000, Sheppard2008}} The barriers are calculated in a conventional spinel cell (32 anions), which ensures a minimum distance of $\sim$~8~\AA{} between the elastic bands and reduces fictitious interactions with periodic images. We verified that migration barriers do not change appreciably ($<$~3\% deviation) when equivalent calculations are performed in larger supercells (see Figure~S2).  Seven images are introduced between the initial and final end points to capture the saddle point and the migration trajectory. All NEB results are based on the PBE functional, without Hubbard \emph{U}.{\cite{Liu2015,Gautam2015}} The migration barriers in spinel-MgIn$_2$S$_4$ are calculated with compensating electrons added as a background charge to ensure charge-neutrality of the structure at non-stoichiometric Mg concentrations.

As migration barriers are calculated in the conventional spinel cell, the degree of inversion ($i$) that can be modeled is constrained by the migration trajectory under consideration, in both the oxide and the sulfide. For example, along Hop 2 (Figure~{\ref{fig:2}}b), a 3/6 Mg ring site occupancy leads to 3 Mn/In atoms in the $8a$ sites, and consequently results in $i \sim 3/8 = 0.375$. Similarly, the barrier calculations along the $16d-48f-16d$ topology (Hops 3, 4, and 5), which require a minimum of 2 Mg atoms in the $16d$ sites (or 2 Mn/In sites in the $8a$), correspond to $i \sim 0.25$. 

Monte-Carlo simulations are used to estimate the Mg percolation thresholds (x$_{crit}$) and the fraction of extractable Mg ions (x$_{ext}$). A 6$\times$6$\times$6 supercell of the primitive spinel structure is used, which corresponds to 1728 anion atoms (Figure~S6 plots convergence behavior with supercell size). In MC simulations, a network of Mg sites is considered percolating when it spans the periodic boundaries of the simulation cell in one or more directions.{\cite{Newman2000}} Inversion in the spinel is introduced during MC sweeps by labelling a number of random $8a$ and $16d$ sites, corresponding to the degree of inversion, as part of the ``Mg sub-lattice''. For example, the Mg sub-lattice in a normal spinel consists of all $8a$ sites. However, in an inverted spinel (with the degree of inversion $i$) the Mg sub-lattice will be composed of (1-$i$)\% of all $8a$ sites and ($i/2$)\% of all $16d$ sites.

To evaluate the M composition at which percolation occurs, a MC sweep is performed with the following steps:{\cite{Newman2000}} ($i$) the supercell is initialized with M atoms in both M and Mg ``sub-lattices'', corresponding to a M$_3$X$_4$ (X~=~O, S) stoichiometry, ($ii$) M atoms on the Mg sub-lattice are randomly changed to Mg, ($iii$) after all Mg sub-lattice sites are changed (i.e., a stoichiometry of MgM$_2$X$_4$ is attained), M atoms on the M sub-lattice are randomly flipped to Mg. During an MC sweep, once a Mg atom replacement results in the formation of a percolating network, the current Mg concentration (x$_{\rm Mg}$) is taken as an estimate of the percolation threshold (x$_{crit}$), while for x~$>$~x$_{crit}$, the fraction of sites within the percolating network, x$_{ext}$, is stored. The values of x$_{crit}$ and x$_{ext}$ are averaged over 2000 MC sweeps to guarantee well-converged estimates. The effect of vacancies on Mg percolation in the Mn-spinel is captured by initializing the Mg sub-lattice with varying vacancy concentrations, at a given degree of inversion, corresponding to a Vac$_{\rm z}$Mn$_{\rm 3-z}$O$_4$ stoichiometry (z~$\leq 1$). Whenever vacancies are initialized in a supercell, only the Mn atoms are changed to Mg during a MC sweep.


\section{Results}
\label{sec:results}
\subsection{MgMn$_2$O$_4$}
\label{sec:results_MnO2}

Figure~{\ref{fig:3}} plots the ranges of Mg$^{2+}$ migration barriers in Mg$_{\rm x}$Mn$_2$O$_4$ (\emph{y}-axis) for all hops of Figure~{\ref{fig:2}} and Table~{\ref{tb:hops}}, while the raw data is included in Figure~S3 of the SI. The migration barriers are calculated with respect to the absolute energies of the end points, nominally identical for a given Mg$^{2+}$ hop. However, there are a few cases where the end point energies are different, since the local symmetry of the cation decoration is broken differently across the end points (e.g., 3/6 hop in Figure~S3b). In such cases, the barrier is reported with respect to the end point with the lowest energy. The dotted black line in Figure~{\ref{fig:3}} is the upper-limit of the Mg migration barrier, as required for reasonable battery performance,{\cite{Canepa2017}} and is used to determine the percolation thresholds (see Section~{\ref{sec:results_percolation}}). For a Mg$_{\rm x}$Mn$_2$O$_4$ cathode particle of size $\sim$~100~nm being (dis)charged at a C/3 rate at 60$^{\circ}$C, the migration barrier upper-limit is $\sim$~750~meV (the upper-limit decreases to $\sim$~660~meV at 298~K).{\cite{Canepa2017}} Since full-cell Mg batteries so far have displayed superior performance at $\sim$~60$^{\circ}$C than at $25^{\circ}$C,{\cite{Aurbach2000,NazarSunDuffortEtAl2016}} the value of $\sim$~750~meV has been used as the cut-off to differentiate ``open'' and ``closed'' Mg$^{2+}$ migration channels. In terms of notations, the fractions used in Hop 2 (e.g., 1/6, 2/6, etc., yellow rectangle in Figure~{\ref{fig:3}}) correspond to the fraction of $16d$ ring sites (Figure~{\ref{fig:2}}b) that are occupied by Mg$^{2+}$. The terms ``8a empty" and ``8a full" along Hops 3, 4, and 5 in Figure~{\ref{fig:3}} indicate that the corner-$8a$ sites (Figures~{\ref{fig:2}}c, d, and e) are vacant and occupied by cations, respectively.  x$_{\rm Mg}$ in Figure~{\ref{fig:3}} is the Mg concentration in the cell used for the barrier estimation, corresponding to the ``dilute Mg" (x$_{\rm Mg} \sim 0$, solid red lines) and ``dilute vacancy" (x$_{\rm Mg} \sim 1$, dashed blue lines) limits.

\begin{figure*}[t!]
\includegraphics[width=\textwidth]{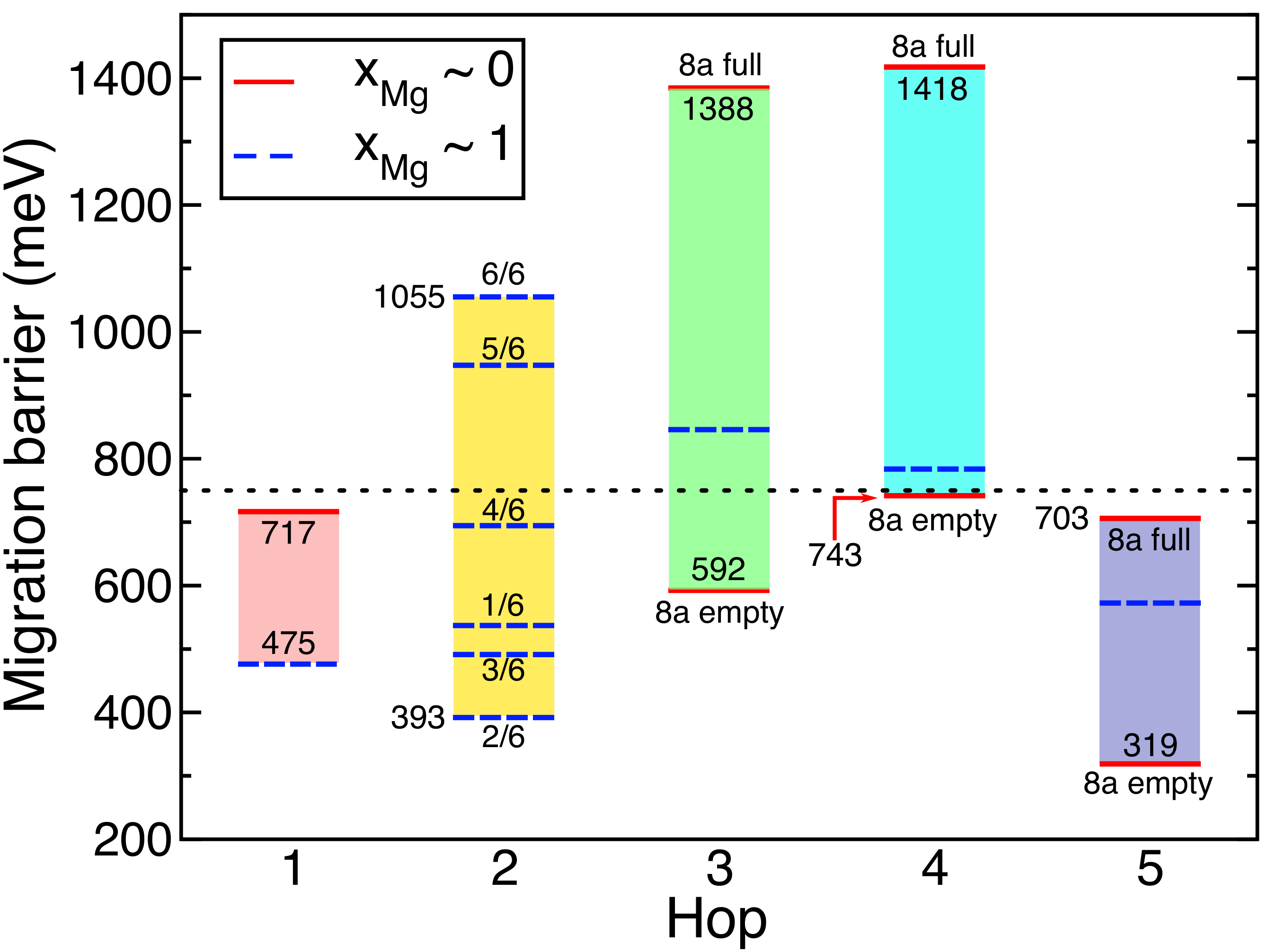}
\caption{ Ranges of Mg$^{2+}$ migration barriers along the hops considered in spinel-Mg$_{\rm x}$Mn$_2$O$_4$. The dotted black line indicates the upper-limit of migration barriers ($\sim$~750~meV) used to distinguish open and closed migration channels in percolation simulations. Solid red and dashed blue lines correspond to dilute Mg (x$_{\rm Mg} \sim 0$) and dilute vacancy (x$_{\rm Mg} \sim 1$) limits. Fractions along Hop 2 indicate the occupancy of Mg$^{2+}$ in the $16d$ ring sites, while the legend ``$8a$ full (empty)'' corresponds to cation-occupied (vacant) corner-$8a$ sites along Hops 3 -- 5. The barriers along Hop 1 are calculated at $i \sim 0$, while Hops 3 -- 5 have been done at $i \sim 0.25$. Along Hop 2, $i$ varies with Mg occupancy of the ring sites, ranging from $i \sim 0.125$ at 1/6 Mg to $i \sim 0.75$ at 6/6 Mg. The raw data from Nudged Elastic Band calculations are displayed in Figure~S3 of the SI.
\label{fig:3} }
\end{figure*}

Mg migration barriers along Hop 1 ($tet-oct-tet$, normal spinel) at the dilute Mg and dilute vacancy limits are $\sim$~717~meV and $\sim$~475~meV, respectively (red rectangle in Figure~{\ref{fig:3}}), in good agreement with previous studies.{\cite{Liu2015,Rong2015,Ling2016}} Note that the dilute Mg (vacancy) limit for Hop 1 corresponds to the regime when no $8a$ sites, other than those required to model the hop, are occupied by Mg (vacancies). Since the migration barriers at both Mg concentration limits are below $\sim$~750~meV, Hop 1 is always open for Mg migration. Barriers along Hop 2 (yellow rectangle in Figure~{\ref{fig:3}}) decrease initially with Mg occupation of the $16d$ ring sites ($\sim$~393~meV at 2/6 vs.\ 536~meV at 1/6) before increasing beyond 750~meV at 5/6 and 6/6 Mg. The non-monotonic variation of the migration barriers along Hop 2 is due to the gradual destabilization of the $16c$ site. The increasing instability of the $16c$ also changes the migration energy profile (Figure~S3b) from ``valley''-like{\cite{Rong2015}} at 1/6 Mg to ``plateau''-like at 5/6 Mg. Figure~S14 shows the Mg migration barriers along Hop 2 when the ring sites are occupied by vacancies instead of Mg$^{2+}$. 

In the case of the $oct-tet-oct$ Hops 3 and 4 (green and cyan rectangles in Figure~{\ref{fig:3}}), which respectively have $tet$ Mg and Mn edge-sharing with the intermediate $48f$ site, the barriers vary drastically based on Mg content and occupancy of the corner-$8a$ sites. For example, at ($i$)  x$_{\rm Mg} \sim 0$ and vacant corner-$8a$, the barrier along Hop 3 ($\sim$~592~meV) is well below the upper-bound of 750~meV, while the barrier is comparable along Hop 4 ($\sim$~743~meV). At ($ii$) x$_{\rm Mg} \sim 0$ and cation-occupied corner-$8a$, the barriers along Hops 3 and 4 increase significantly ($\sim$~1388~meV and $\sim$~1418~meV) and surpass the upper-limit set for open channels. Eventually, at ($iii$) x$_{\rm Mg} \sim 1$ (cation-occupied corner-$8a$), the barriers decrease to $\sim$~845~meV and $\sim$~784~meV along Hops 3 and 4, respectively. Note that the barriers along Hops 3 and 4 in Figure~{\ref{fig:3}} are calculated at a degree of inversion, $i \sim 0.25$. At a higher degree of inversion ($i \sim 1$) and x$_{\rm Mg} \sim 1$ (cation-occupied corner-$8a$), the barrier is $\sim$~1039~meV along Hop 4 (Figure~S5). Hence, from the data of Figure~{\ref{fig:3}}, Hop 3 is considered closed for Mg migration whenever the corner-$8a$ sites are cation-occupied, while Hop 4 is always considered a closed channel.

Mg migration barriers decrease significantly if the edge-$8a$ is vacant (i.e., along Hop 5). For example, the migration barriers along Hop 5 (purple rectangle in Figure~{\ref{fig:3}}) are well below that of Hops 3 and 4 across the scenarios of ($i$) low Mg, vacant corner-$8a$ (319~meV for Hop 5 vs. 592 and 743~meV for Hops 3 and 4, respectively), ($ii$) low Mg, cation-occupied corner-$8a$ (703~meV vs.\ 1388 and 1418~meV), and ($iii$) high Mg, cation-occupied corner-$8a$ (570~meV vs. 845 and 784~meV). Hence, Hop 5 is always open for Mg migration, since the barriers are below the upper limit of 750~meV.

In summary, the $tet-oct-tet$ pathway (Hops 1 and 2) remains open for Mg migration in MgMn$_2$O$_4$ until a high degree of Mg occupation on the $16d$ ring sites (i.e., $\geq 5/6$~Mg) is present, which corresponds to high degrees of inversion ($i > 0.625$). The $oct-tet-oct$ pathway is open only when the edge-$8a$ is vacant (Hop 5) or when the corner-$8a$ are vacant with Mg in the edge-$8a$ (Hop 3). 

\subsection{MgIn$_2$S$_4$}
\label{sec:results_InS2}
Figure~{\ref{fig:4}} plots the Mg$^{2+}$ migration barriers in MgIn$_2$S$_4$ for the hops of Figure~{\ref{fig:2}} (the raw data are shown in Figure~S4). Since we consider MgIn$_2$S$_4$ as an ionic conductor, off-stoichiometric Mg concentrations are not of interest. Hence, all hops in Figure~{\ref{fig:4}} are evaluated at the dilute vacancy limit (x$_{\rm Mg} \sim 1$, dashed blue lines in Figure~{\ref{fig:4}}).  The fractions used (1/6, 2/6, etc.) in Figure~{\ref{fig:4}} are the number of $16d$ ring sites occupied by Mg$^{2+}$ in Hop 2. Along Hops 3 -- 5, we use cation-occupied corner-$8a$ sites (i.e., ``$8a$ full'' in Figures~S4c, d, and e). The upper-limit of the Mg migration barrier for classifying open and closed migration channels (as indicated by the dotted black line in Figure~{\ref{fig:4}}) is set to $\sim$~500~meV, based on migration barriers of $\sim$~400~--~500~meV observed in fast Li-ion conductors, such as Garnets and Si-based thio-LISICONs.{\cite{BachmanMuyGrimaudEtAl2016}}

\begin{figure*}[t!]
\includegraphics[width=\textwidth]{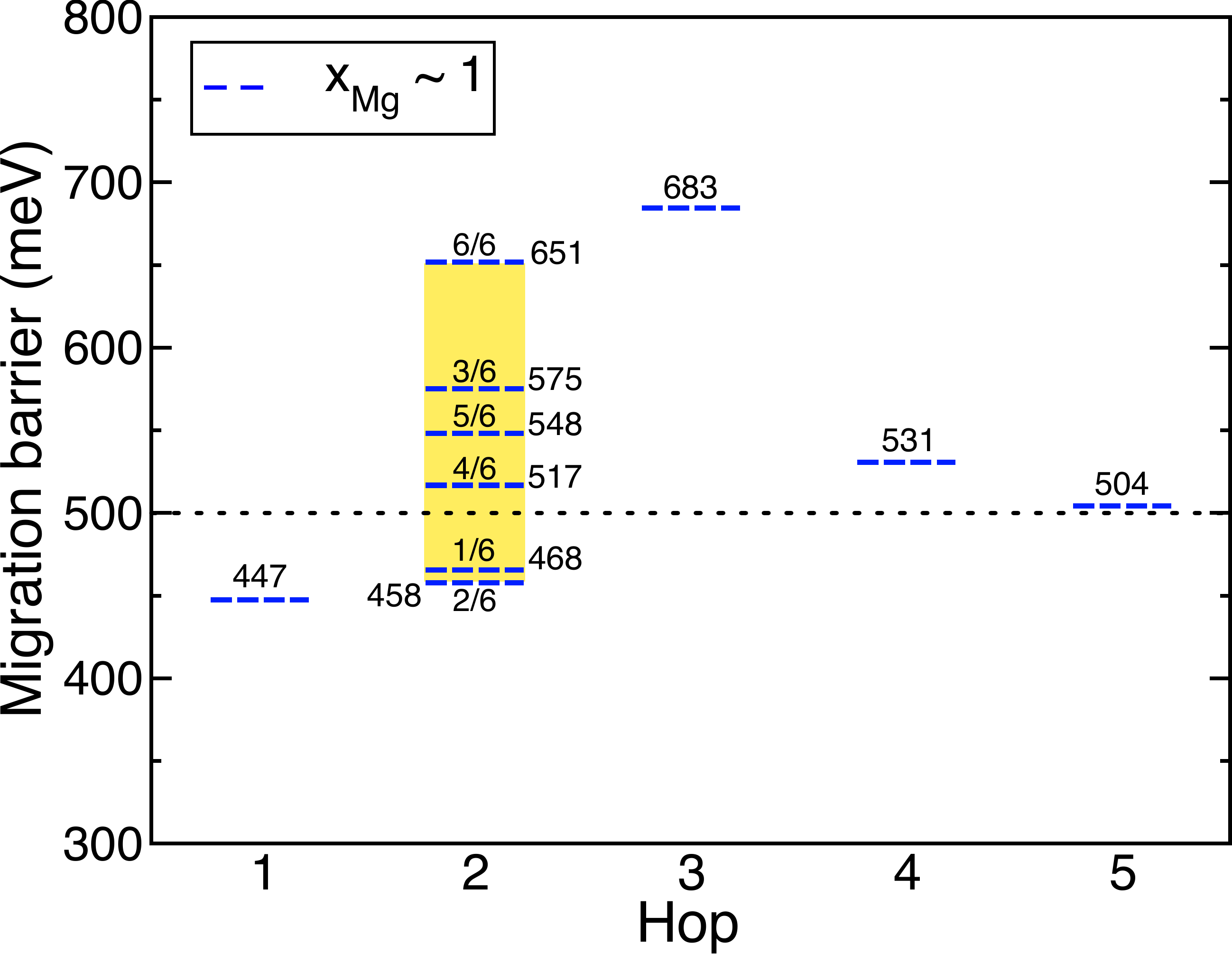}
\caption{
Mg$^{2+}$ migration barriers along each possible hop in spinel-MgIn$_2$S$_4$. The dotted black line indicates the upper-limit of migration barriers ($\sim$~500~meV) used to distinguish open and closed migration channels in percolation simulations. Dashed blue lines indicate the dilute vacancy (x$_{\rm Mg} \sim 1$) limit. Fractions along Hop 2 indicate the occupancy of Mg$^{2+}$ in the $16d$ ring sites, while the corner-$8a$ sites are cation-occupied across Hops 3 -- 5. The barrier along Hop 1 is calculated at $i \sim 0$, while Hops 3 -- 5 have been done at $i \sim 0.25$. Along Hop 2, $i$ varies with Mg occupancy of the ring sites, ranging from $i \sim 0.125$ at 1/6 Mg to $i \sim 0.75$ at 6/6 Mg. The raw data from Nudged Elastic Band calculations are displayed in Figure~S4.
\label{fig:4}
}
\end{figure*}

In the case of Hop 1, the barrier is $\sim$~447~meV, well below the upper limit of $\sim$~500~meV. Mg migration barriers along Hop 2 (yellow rectangle in Figure~{\ref{fig:4}}) follow trends similar to that of MgMn$_2$O$_4$ (Figure~{\ref{fig:3}}). For example, at low Mg occupation of the ring sites (1/6 or 2/6 Mg), the barrier is below the limits for percolating diffusion, before increasing beyond 500~meV at higher Mg content in the ring sites ($>$ 3/6 Mg). Also, the shape of the migration energy curve changes from a ``valley" at 1/6 Mg (solid black line in Figure~S4b) to a ``plateau" beyond 2/6 Mg (solid red line in Figure~S4b), indicating that the $16c$ site becomes progressively unstable with increasing Mg occupation of the ring $16d$.

Along the $16d-48f-16d$ pathways (Hops 3, 4 and 5), the migration barriers are always higher than 500 meV, irrespective of the occupancy of the edge-$8a$. Indeed, the magnitude of the barriers are $\sim$~683~meV, $\sim$~531~meV, and $\sim$~504~meV for Mg-occupied, In-occupied and vacant edge-$8a$, respectively, indicating that the $oct-tet-oct$ pathway will not be open for Mg$^{2+}$ migration.

\subsection{Percolation thresholds}
\label{sec:results_percolation}
Based on the data of Figures~{\ref{fig:3}} and {\ref{fig:4}}, and the upper limits of Mg migration barriers set for MgMn$_2$O$_4$ (750~meV) and MgIn$_2$S$_4$ (500~meV), we compiled a list of conditions that enable the opening of the possible hops in Table~{\ref{tb:percolation}}. For example, Hop 1 ($8a-8a$) is open for all values of x$_{\rm Mg}$ and $i$ for both Mg$_{\rm x}$Mn$_2$O$_4$ and MgIn$_2$S$_4$. Both the oxide and the sulfide spinel exhibit high barriers ($> 1$~eV) for a $16d-8a$ hop (Figure S8), which would limit Mg transfer between an octahedral $16d$ site and an adjacent tetrahedral $8a$ site. Thus, in our percolation simulations, the $8a-8a$ (Hops 1 and 2) and the $16d-16d$ (Hops 3, 4, and 5) channels remain decoupled, and a percolating network consists solely of either $8a-8a$ or $16d-16d$ channels.

\begin{table}[th]
\caption{\label{tb:percolation} Summary of rules used during percolation simulations with the conditions for an open channel. The upper limit of migration barriers used to distinguish between open and closed channels is 750~meV and 500~meV for MgMn$_2$O$_4$ and MgIn$_2$S$_4$, respectively.}
%
\begin{tabular*}{\columnwidth}{@{\extracolsep{\fill}}lcl@{}}
\hline\hline
Hop & Topology & Open under condition \\
\hline
\multicolumn{3}{c}{\textbf{MgMn$_2$O$_4$ -- 750 meV}} \\
\hline
1 & $8a-16c-8a$ & Always open \\
2 & $8a-16c-8a$ & Max.~4/6 ring sites with Mg \\
3 & $16d-48f-16d$  & Corner $8a$ vacant \\
4 & $16d-48f-16d$  & Always closed \\
5 & $16d-48f-16d$  & Always open \\
\hline
\multicolumn{3}{c}{\textbf{MgIn$_2$S$_4$ -- 500 meV}} \\
\hline
1 & $8a-16c-8a$ & Always open \\
2 & $8a-16c-8a$ & Max.~2/6 ring sites with Mg \\
3 & $16d-48f-16d$  & Always closed \\
4 & $16d-48f-16d$  & Always closed \\
5 & $16d-48f-16d$  & Always closed \\
\hline
\hline
\end{tabular*}
\end{table}

Figures~{\ref{fig:5}}a and b plot the percolation threshold (x$_{crit}$, black lines), at various degrees of inversion ($i$) in Mn$_{\rm 3-x}$O$_4$ and In$_{\rm 3-x}$S$_4$. The dashed yellow lines indicate the stoichiometric spinel, i.e., M:X~=~2:4. The blue (red) shaded region corresponds to Mg concentration ranges which do (do not) exhibit percolation. The x-axis in Figure~{\ref{fig:5}} begins at a M$_3$X$_4$ (i.e., 50\% M-excess or 100\% Mg-deficient) configuration and spans concentrations up to Mg$_{1.5}$M$_{1.5}$X$_4$ (i.e., 25\% M-deficient, 50\% Mg-excess). Generally, percolation thresholds in the M-excess domain (i.e., x$_{crit} < 1$) are desirable as this implies that the stoichiometric spinel will possess percolating networks and will facilitate macroscopic Mg transport.

In the case of cathodes (Mn$_2$O$_4$), Mg deintercalation from the framework creates vacancies, which can facilitate the formation of Mg percolating networks by opening certain migration channels (e.g., Hop 5 in MgMn$_2$O$_4$, Figure~{\ref{fig:3}}). Therefore, we explored the variation of the percolation threshold with vacancy concentration (``z'' in Figure~{\ref{fig:5}}a) in the Mn-spinel. 
For the sake of simplicity, x in Figure~{\ref{fig:5}} refers to the sum of Mg and vacancy concentrations. For example, x~=~1 and z~=~0.5 (green circle on the dashed black line) in Figure~{\ref{fig:5}}a indicates a composition of Mg$_{0.5}$Vac$_{0.5}$Mn$_2$O$_4$, while x~=~0.6 and z~=~0 (green square) corresponds to the M-excess spinel-Mg$_{0.6}$Vac$_{0}$Mn$_{2.4}$O$_4$. 

The percolation threshold in the absence of vacancies (z~=~0) is indicated by the solid black line in Figure~{\ref{fig:5}}a. When vacancies are introduced, the threshold decreases, as indicated by the x$_{crit}$ at z~=~0.4 (dotted black line) or 0.5 (dashed) consistently exhibiting lower values than x$_{crit}$ at z~=~0 in Figure~{\ref{fig:5}}a. For example, at x~=~0.8 and $i = 0.5$ (indicated by the green star in Figure~{\ref{fig:5}}a), the spinel does not form a percolating network when there are no vacancies (z~=~0, Mg$_{0.8}$Mn$_{2.2}$O$_4$), since x$_{crit} \sim 0.88 > 0.8$. However, the structure can percolate Mg when vacancies are introduced (z~=~0.5, Mg$_{0.3}$Vac$_{0.5}$Mn$_{2.2}$O$_4$), as x$_{crit}$ reduces to $\sim 0.52 < 0.8$. In a case such as this, the initial cathode structure may not be percolating, but introducing vacancies in the initial part of the charge can create a percolating zone on the cathode particle surface through which further Mg-removal can occur. However, upon discharge the percolating structure could easily become non-percolating if polarization increases the surface Mg concentration too rapidly.

At any degree of inversion, the magnitude of x$_{crit}$ varies non-monotonically and reduces only up to a vacancy content, z~=~0.4 or 0.5 (see Figure~S7a). Indeed, at x~=~0.8 and $i = 0.5$ (green star), an increase in z beyond 0.5 (such as z~=~0.6, Mg$_{0.2}$Vac$_{0.6}$Mn$_{2.2}$O$_4$), causes the x$_{crit}$ to increase to $\sim 0.6$, but the spinel continues to percolate. Thus, the shaded grey region in Figure~{\ref{fig:5}}a, which is bound by the z~=~0.4, 0.5 and 0 lines represents the extent of variation of x$_{crit}$ with vacancy content in the cathode. Notably, the lowest value of x$_{crit}$ is obtained at z~=~0.4 for $0 \leq i \leq 0.35$ and $0.595 \leq i < 0.77$, and at z~=~0.5 for $0.35 \leq i \leq 0.595$, respectively. 

\begin{figure*}[t!]
\centering
\includegraphics[scale=0.55]{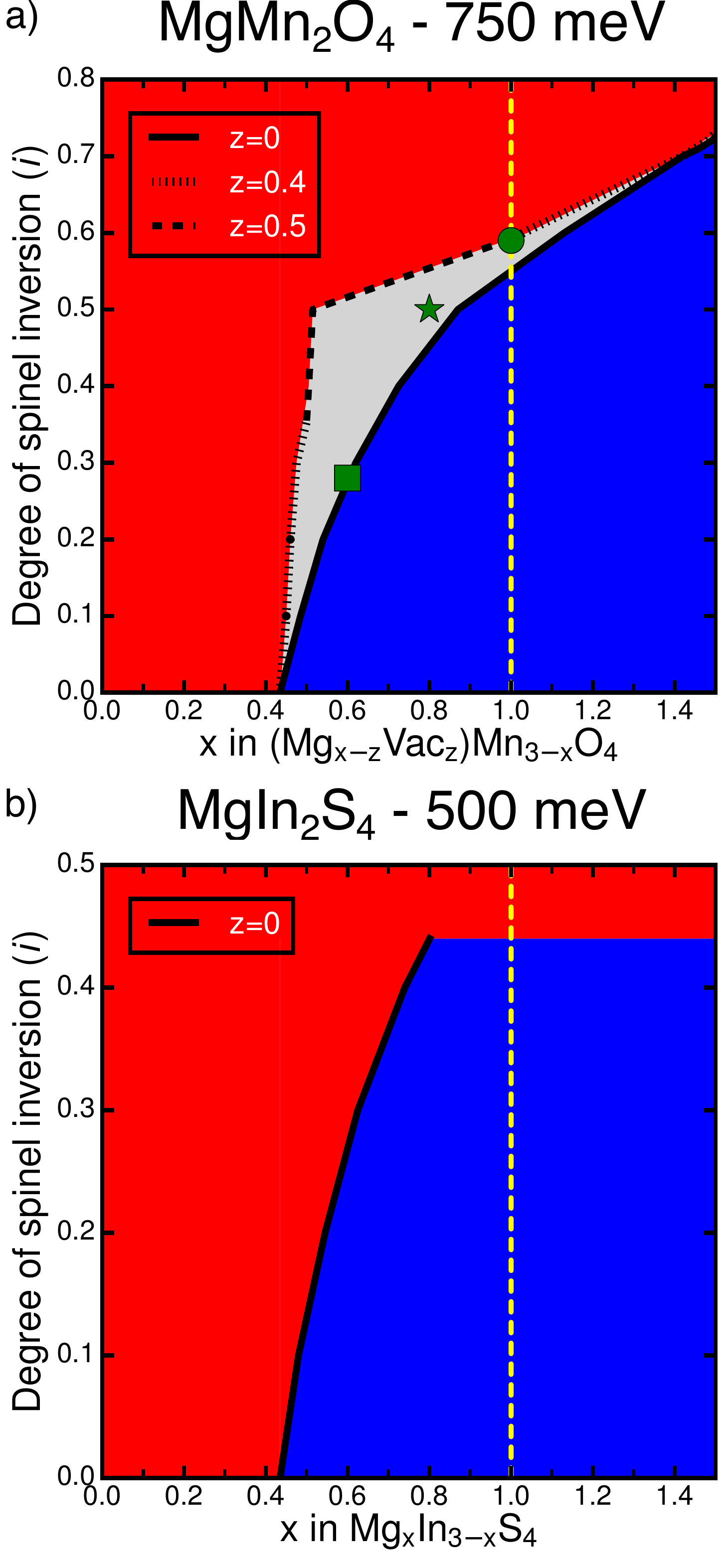}
\caption{
The critical concentration for Mg percolation (x$_{crit}$) in the (Mg$_{\rm x-z}$Vac$_{\rm z}$)Mn$_{\rm 3-x}$O$_4$ (a) and Mg$_{\rm x}$In$_{\rm 3-x}$S$_4$ (b) spinels are plotted as black lines at different degrees of spinel inversion $i$. The stoichiometric spinel concentration (M:X~=~2:4) is indicated by the dashed yellow lines. Note that the zero on the $x$-axis corresponds to a stoichiometry of M$_3$X$_4$ (M = Mn/In and X = O/S). z indicates the vacancy concentration in the structure. The shaded red (blue) region in both panels indicates the Mg concentration range where macroscopic Mg diffusion is not possible (possible). The shaded grey region in panel (a) refers to the range of variation of the percolation threshold with vacancy content in the oxide cathode. The green circle, square and star in panel (a) correspond to sample scenarios discussed in the text.
\label{fig:5}
}
\end{figure*}

The stoichiometric \{Mg/Vac\}Mn$_2$O$_4$ spinel at $i = 0$ (dashed yellow line in Figure~{\ref{fig:5}}a), permits macroscopic Mg diffusion, since the percolation threshold (x$_{crit} \sim 0.44$ for z~=~0~--~0.4) is in the Mn-excess domain (i.e., x$_{crit} < 1$). When vacancies are absent in the stoichiometric spinel (z~=~0), which corresponds to the discharged MgMn$_2$O$_4$ composition, the structure percolates Mg up to $i \sim 0.55$. Upon charging, the presence of vacancies (z~=~0.5) enables Mg percolation within Mg$_{0.5}$Vac$_{0.5}$Mn$_2$O$_4$ up to $i \sim 0.59$.  At higher degrees of inversion ($0.59 < i < 0.77$), the oxide spinel requires Mn-deficient concentrations (i.e., x~$> 1$) to facilitate Mg percolation, as illustrated by x$_{crit} \sim 1.05 - 1.13$ (z~=~0.5 -- 0) at $i~=~0.6$. At $i > 0.77$, the oxide does not form a percolating Mg network at any level of Mn-deficiency (for z~$\leq 1$) in the lattice. 

In stoichiometric ionic conductors, such as MgIn$_2$S$_4$, the vacancy concentration is low and therefore vacancies are not expected to play a major role in macroscopic Mg transport. Specifically in MgIn$_2$S$_4$, vacancies do not open additional migration channels, as indicated by the closed Hop~5 in Figure~{\ref{fig:4}}.  Indeed, the percolation threshold in the In-spinel does not change up to a vacancy content, z~=~0.2 in the structure (see Figure~S7b). At z~=~0, the x$_{crit}$ in In$_{\rm 3-x}$S$_4$ (solid black line in Figure~{\ref{fig:5}}b) increases continuously with increase in inversion, with x$_{crit} \sim 0.435$, and 0.74 at $i = 0$, and 0.4, respectively. Thus, at low $i$, stoichiometric MgIn$_2$S$_4$ should exhibit significant ionic conductivity. However, at higher degrees of inversion ($i > 0.44$), the sulfide spinel does not form percolating networks at any Mg-concentration, owing to the absence of open $16d-16d$ channels in combination with the $8a-8a$ channels being closed beyond 2/6 Mg ring site occupancy (Table~{\ref{tb:percolation}}). 

In general, mobility requirements in an ionic conductor are more stringent than in a cathode, consistent with the stricter cut-off of 500~meV we applied to the migration barriers in MgIn$_2$S$_4$.{\cite{WangRichardsOngEtAl2015,BachmanMuyGrimaudEtAl2016}} Indeed, a sulfide spinel Mg-cathode (such as Mg$_{\rm x}$Ti$_2$S$_4${\cite{NazarSunDuffortEtAl2016}}) exhibiting similar activation barriers with inversion as MgIn$_2$S$_4$ will not suffer from any percolation bottlenecks, since the barriers across all cation arrangements are well below the milder 750~meV cut-off set for cathodes (Figure~{\ref{fig:4}}).

\subsection{Impact of inversion on cathode electrochemistry}
\label{sec:voltage}
Under ideal conditions, the structure of an ionic conductor (such as MgIn$_2$S$_4$) should not undergo significant changes during operation. Thus, the extent of inversion should, in principle, be measured using characterization experiments post-synthesis (the calculated formation energies of various inverted configurations in spinel-MgIn$_2$S$_4$ are plotted in Figure~S11). However in a cathode material such as Mg$_{\rm x}$Mn$_2$O$_4$, which can generate mobile Mn$^{2+}$ ions (Figure S9) through disproportionation of Mn$^{3+}$, the degree of inversion ($i$) can change during electrochemical cycling.{\cite{Reed2001,Reed2004}} Consequently, structural changes in a cathode during cycling should manifest themselves as changes in the voltage profile and observed capacity, which can be benchmarked with theoretical predictions.{\cite{Canepa2017,SaiGautam2015}} To evaluate the effect of inversion on the  voltage profile of Mg$_{\rm x}$Mn$_2$O$_4$, we calculated the phase diagram and energy of the intercalation system at 0~K as a function of Mg content under various degrees of inversion.{\cite{Aydinol1997,Anton2000_SpinelMnO2,Reed2001,SaiGautam2015}}

To evaluate the ground state hull of the Mg$_{\rm x}$Mn$_2$O$_4$ system, we enumerated over 400 Mg-vacancy configurations, at different Mg concentrations (x$_{\rm Mg}$ = 0, 0.25, 0.5, 0.75 and 1) and different degrees of inversion ($i$= 0, 0.25, 0.5, 0.75 and 1).  Figure~{\ref{fig:6}}a displays structures with formation energies ($y$-axis) below 200~meV/Mn$_2$O$_4$ at different Mg concentrations ($x$-axis), and the formation energies of all the Mg-vacancy configurations considered are plotted in Figure~S10 of the SI. Notably, formation energies in Figure~{\ref{fig:6}}a have been referenced to the non-inverted ($i$~=~0), empty Mn$_2$O$_4$ and magnesiated (MgMn$_2$O$_4$) spinel configurations. For each configuration, the degree of inversion is indicated by the corresponding symbol used, ranging from  $i = 0$ (black circles) to $i = 1$ (red stars).

Overall, the Mg$_{\rm x}$Mn$_2$O$_4$ system is phase separating at 0~K across non-inverted ($i = 0$) MgMn$_2$O$_4$ and Mn$_2$O$_4$ domains, since the ground state hull of the system (dashed black line in Figure~{\ref{fig:6}}a) only exhibits two configurations (i.e., MgMn$_2$O$_4$ and Mn$_2$O$_4$). Some solubility at low Mg content may be possible given the low positive mixing energy at x$_{\rm Mg} = 0.25$ for the non-inverted spinel ($E_{formation} \sim 14$~meV/Mn$_2$O$_4$). At higher Mg content, the formation energies are very high for the non-inverted spinel (Figure~S10), making a solid solution behavior very unlikely. Inversion becomes likely to occur at intermediate Mg compositions, as the low positive formation energies are on the scale of the configurational entropy. For example, $E_{formation} \sim$~11~meV/Mn$_2$O$_4$ at $i$~=~0.25 and x$_{\rm Mg} = 0.5$ (green square at x~=~0.5 in Figure~{\ref{fig:6}}a). Hence, inversion at intermediate states of magnesiation is likely. While Mg by definition has to be mobile in Mn$_2$O$_4$ to operate as a cathode, Mn mobility, which is required for spinel inversion to occur, depends strongly on its valence state.{\cite{Reed2001,Reed2004}} Typically, Mn$^{3+}$ can be mobile through a temporary disproportionation mechanism, generating mobile Mn$^{2+}$ (Figure~S9).{\cite{Reed2001,Reed2004}}

\begin{figure*}[t!]
\centering
\includegraphics[scale=0.56]{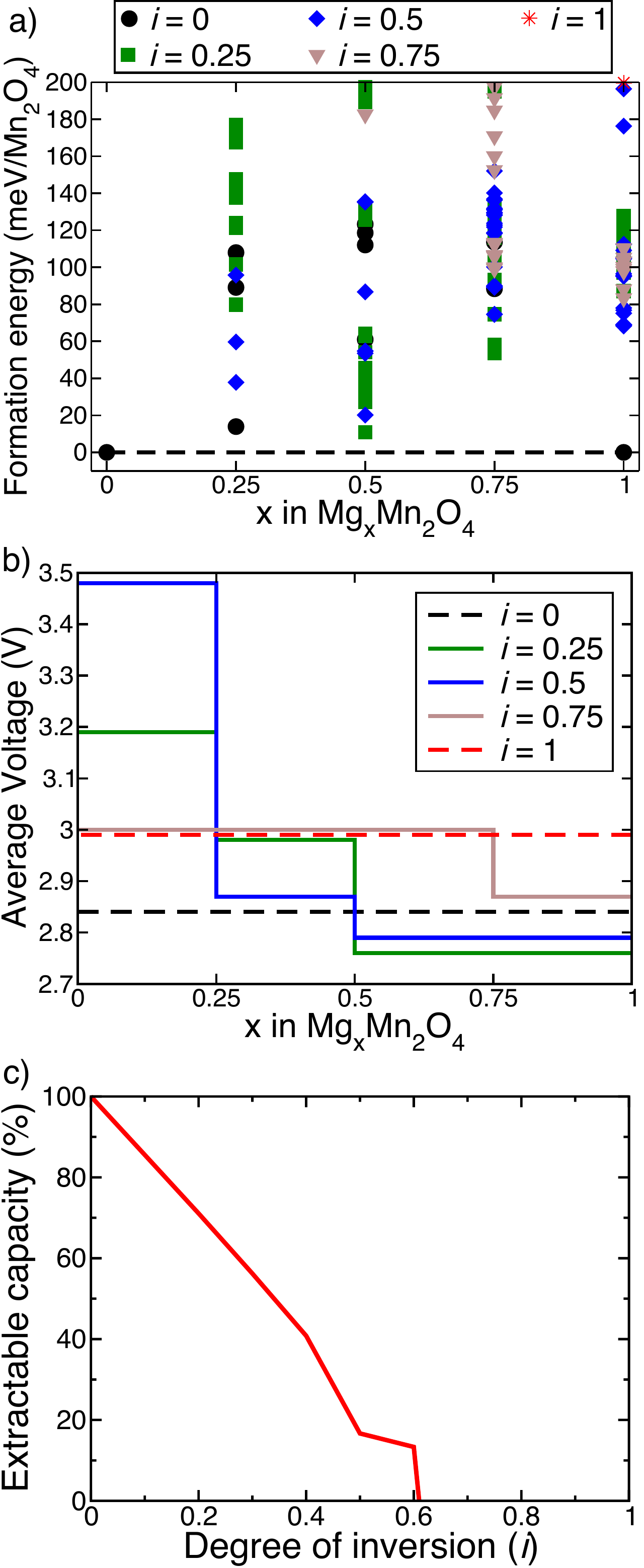}
\caption{
(a) Ground state hull (or 0~K phase diagram) of the Mg$_{\rm x}$Mn$_2$O$_4$ system, with the zero of the formation energy referenced to the non-inverted ($i$=0) magnesiated (MgMn$_2$O$_4$) and empty (Mn$_2$O$_4$) spinel configurations. (b) Average voltage curves under $i$ in Mg$_{\rm x}$Mn$_2$O$_4$, obtained using the lowest formation energy structures at each $i$ across Mg concentrations. (c) The percentage of the theoretical capacity that can be reversibly extracted is plotted as a function of inversion in stoichiometric MgMn$_2$O$_4$.
\label{fig:6}
}
\end{figure*}

Figure~{\ref{fig:6}}b plots the average voltages as a function of x$_{\rm Mg}$ at different $i$ by taking the lowest $E_{formation}$ configuration at each $i$ and x$_{\rm Mg}$.{\cite{Aydinol1997}} The average voltage for Mg insertion in the non-inverted ($i$~=~0) configuration is $\sim$~2.84~V (dashed black line in Figure~{\ref{fig:6}}b), in agreement with previous theoretical estimates.{\cite{Liu2015,Jain2013}} Inversion does increase the average insertion voltage (averaged over x$_{\rm Mg} = 0$ to 1) marginally compared to the normal spinel, with specific values of $\sim$~2.92, 2.99, 2.97 and 2.99 V at $i$~=~0.25, 0.5, 0.75 and 1, respectively. Notably, the phase behavior of the Mg$_{\rm x}$Mn$_2$O$_4$ system under inversion will be different compared to the normal spinel due to the formation of metastable inverted states at intermediate Mg compositions. 

The extractable Mg content (x$_{ext}$, see Section~{\ref{sec:percolation}}), obtained as a function of inversion from our Monte-Carlo simulations, indicates the extractable capacity of a cathode particle, and is shown in Figure~{\ref{fig:6}}c for stoichiometric MgMn$_2$O$_4$. The $y$-axis indicates the \% of the cathode's theoretical capacity ($\sim$~270~mAh/g for MgMn$_2$O$_4$), that can be cycled reversibly. At low degrees of inversion, the extractable capacity in the stoichiometric spinel decreases roughly linearly with the degree of inversion, reaching $\sim$~41\% ($\sim$~110~mAh/g) at $i = 0.4$. The extractable Mg content decreases more rapidly from $i = 0.4$ to $i = 0.5$, before stabilizing around $\sim$~15\% ($\sim$~40~mAh/g) between $i = 0.5$ and 0.6. Eventually, none of the Mg becomes extractable beyond $i = 0.61$, reflecting the trends in the percolation thresholds (x$_{crit} \sim 0.59$ at stoichiometric MgMn$_2$O$_4$, Figure~{\ref{fig:5}}a) at high degrees of inversion. Note that, the overall amount of cyclable Mg from a cathode particle is influenced both by the extractable Mg (shown in Figure~{\ref{fig:6}}c) and by the phase behavior as a function of x$_{\rm Mg}$. For example, if the Mg removal occurs via a two-phase reaction (as is the case for the non-inverted spinel), then the presence of a  non-percolating layer on the surface may prevent extraction of Mg from the bulk, even if percolation conditions are still favorable in the bulk material.


\section{Discussion}
\label{sec:discussion}
In this work, we have used DFT-based NEB calculations to assess the changes in the activation barrier for Mg$^{2+}$ migration arising from inversion in both oxide (MgMn$_2$O$_4$) and sulfide (MgIn$_2$S$_4$) structures. From our results (Figures~{\ref{fig:3}} and {\ref{fig:4}}), we can conclude that inversion has a significant impact on both oxides and sulfides, by opening and closing specific migration trajectories. In order to extrapolate the impact of the various Mg$^{2+}$ migration barriers on macroscopic Mg diffusion, we estimated the percolation thresholds under different degrees of spinel inversion. Furthermore, we analyzed the impact of spinel inversion on cathode properties of Mg$_{\rm x}$Mn$_2$O$_4$ by evaluating the average voltages and practical capacities at different degrees of inversion.

\subsection{Factors influencing barriers in MgMn$_2$O$_4$}
Trends from activation barriers of Figure~{\ref{fig:3}} suggest that Mg migration along the $8a-16c-8a$ pathways (Hops 1 and 2) can improve significantly with Mg occupation of the $16d$ ring sites (up to 4/6 Mg), at low degrees of inversion. Additionally, the $16d-48f-16d$ channels open for Mg migration whenever the edge-$8a$ is vacant. However, high degrees of inversion detrimentally affect Mg$^{2+}$ motion, due to the closing of both $16d-16d$ (corner- and edge-$8a$ become occupied by the metal cation) and $8a-8a$ channels (high migration barriers at high Mg in the ring sites). Although we have specifically considered the case of spinel-Mg$_{\rm x}$Mn$_2$O$_4$, similar trends can be expected for other oxide spinels, given the similarity in Mg migration barriers along Hop 1 with different $3d$-metals.{\cite{Liu2015}}

Previous studies have used electrostatic considerations to partially explain trends in Li$^+$ activation barriers in a Mn$_2$O$_4$ spinel.{\cite{Xu2010a}} Indeed, the reduction in Mg migration barriers along Hops 1, 3, 4, and 5 (Figure~{\ref{fig:3}}) with increasing Mg concentration can be attributed to lower electrostatic repulsions at the corresponding intermediate sites caused by the reduction of Mn$^{4+}$ to Mn$^{3+}$. For example, the barrier reduces from 717 to 475 meV along Hop 1 and 1388 to 845~meV along Hop 3, as x$_{\rm Mg}$ increases from $\sim$~0 to $\sim$~1. However, Mg$^{2+}$ activation barriers generally depend on steric and bonding constraints in addition to electrostatics, which are often difficult to deconvolute over a range of NEB calculations. For example, the Mg$^{2+}$ activation barriers across Hop 2 (yellow bar in Figure~{\ref{fig:3}}) at low Mg occupation in the ring sites (1/6, 2/6) are lower than Hop 1 (red bar, Figure~{\ref{fig:3}}), which may be attributed to reduced electrostatic repulsion on the intermediate $16c$ (due to Mg$^{2+}$ replacing higher valent Mn in the ring sites). However, barriers along Hop 2 increase beyond Hop 1 and eventually beyond the limit of $\sim$~750~meV at higher Mg in the ring sites (5/6, 6/6), despite lower electrostatic repulsion. Thus, the high Mg content in the ring sites decreases the stability of the intermediate $16c$. One possible reason for the instability of the $16c$ site could arise from charge-deficient oxygen atoms being shared with adjacent, Mg$^{2+}$-occupied (instead of Mn$^{3+/4+}$) $16d$ sites. Indeed, the instability of the $16c$ (e.g., in the case of 6/6 Mg in Hop 2) is quantified by longer (DFT-based) $\sim$~2.3~\AA{} Mg--O bonds, compared to $\sim$~2.08~\AA{} in $16d$ with Mg (along the same hop) and $\sim$~2.13~\AA{} in rocksalt MgO.{\cite{Jain2013}} 

For the $16d-48f-16d$ hops in Figure~{\ref{fig:3}} (Hops 3--5), electrostatic effects are more dominant than for the $tet-oct-tet$ hops (Hops 1, 2), primarily due to the intermediate $48f$ edge-sharing with an $8a$. Indeed, the cation centers in edge-sharing tetrahedra are closer ($\sim$~2.15~\AA{} experimentally between $48f$ and $8a$ in an ideal LiMn$_2$O$_4$-spinel{\cite{Bagci2014}}) than in edge-sharing octahedra ($\sim$~2.88~\AA{} between $16c$ and $16d$). Consequently, the Mg barriers are consistently lower with a vacant edge-$8a$ (Hop 5, Figure~{\ref{fig:3}}) compared to Mg/Mn-filled edge-$8a$ (Hops 3, 4 in Figure~{\ref{fig:3}}). Also, Mg$^{2+}$ activation barriers (at x$_{\rm Mg} \sim 0$) increase significantly when the corner-$8a$ sites are cation-occupied rather than vacant (Figure~{\ref{fig:3}}). A closer look at the cation-cation distances across corner-sharing $48f$ and $8a$ ($\sim$~2.88~\AA{} in ideal LiMn$_2$O$_4$) reveals that the corner-sharing tetrahedra within a spinel framework may experience electrostatic repulsion as high as edge-shared octahedra (i.e., $16c$ and $16d$). Thus, the combination of cation-cation repulsion arising from both edge- and corner-$8a$ sites results in the high barriers along Hops 3 and 4.

\subsection{Barriers in sulfides vs.\ oxides}
Activation barriers calculated in MgIn$_2$S$_4$ (Figure~{\ref{fig:4}}) exhibit similar trends to MgMn$_2$O$_4$ (Figure~{\ref{fig:3}}), resulting from analogous trends in electrostatics, steric and bonding environments. However, the absolute changes in barriers in the sulfide are remarkably lower than the oxide. For example, the absolute difference between the lowest and the highest Mg migration barriers of MgMn$_2$O$_4$ (at x$_{\rm Mg} \sim 1$) across Hops 1 through 5 is $\sim$~662~meV (1055 -- 393 meV), while this is a much lower $\sim$~236~meV (683 -- 447~meV) for MgIn$_2$S$_4$. Similarly, the barriers along the $16d-48f-16d$ trajectory are far less sensitive to the edge-$8a$ occupancy in the sulfide (504--673~meV) than in the oxide  (570--845~meV at x$_{\rm Mg} \sim 1$).  Surprisingly, the migration barrier with an edge-$8a$ occupied by Mg$^{2+}$ is higher ($\sim$~683~meV) than when the edge-$8a$ is occupied by In$^{3+}$ ($\sim$~531~meV), suggesting that the In--S bonding environment screens the higher In$^{3+}$ charge better than the Mg--S bonds screen Mg$^{2+}$.

Lower activation barriers for Mg in sulfides have been reported before,{\cite{Liu2015,LiuJainQuEtAl2016,Canepa2017a}} which have been assigned to robust electrostatic screening, high polarizability, higher degree of covalency and large volume per anion of  S$^{2-}$ compared to O$^{2-}$.{\cite{Canepa2017,WangRichardsOngEtAl2015}} For example, a Mg$_{\rm x}$Ti$_2$S$_4${\cite{NazarSunDuffortEtAl2016}} cathode will not suffer from any percolation bottlenecks, if the barriers across all cation arrangements are similar to the calculated values in MgIn$_2$S$_4$ (i.e., $< 750$~meV, Figure~{\ref{fig:4}}). But a more stringent upper-bound of $\sim$~500~meV on the barrier in a solid-state conductor{\cite{WangRichardsOngEtAl2015,BachmanMuyGrimaudEtAl2016}} indicates that inversion can significantly affect a sulfide ionic conductor by closing all $16d-16d$ channels and several $8a-8a$ channels with high Mg in the $16d$ ring (Figure~{\ref{fig:4}}). Since ionic mobility is expected to improve with larger anions and higher covalency (such as Se$^{2-}$ compared to S$^{2-}$ and O$^{2-}$), inversion is expected to affect Mg-mobility to a lesser extent in Mg-containing Se-spinels, such as MgSc$_2$Se$_4$, compared to oxides and sulfides.

\subsection{Percolation under inversion}
Estimations of percolation thresholds (x$_{crit}$) in the Mg$_{\rm x}$Mn$_{\rm 3-x}$O$_4$ system (Figure~{\ref{fig:5}}a) indicate that spinel inversion should not detrimentally affect macroscopic Mg$^{2+}$ diffusion across the structure up to a fairly high degree of inversion, $i \sim 0.55 - 0.59$. However, Mg-excess concentrations are required to ensure percolating networks form at $i = 0.6 - 0.7$, while the spinel completely ceases to percolate Mg beyond $i = 0.77$ (Figure~{\ref{fig:5}}a). Given the preponderance of conversion reactions under Mg-excess concentrations in the oxide spinel, specifically the decomposition of Mg$_{\rm x}$Mn$_{\rm 3-x}$O$_4$ (x$~>~$1) into MgO and MnO,{\cite{Canepa2017}} it is of paramount importance that the chemically synthesizable, stoichiometric \{Mg/Vac\}Mn$_2$O$_4$ remains percolating. Efforts should be made to reduce or precisely control the amount of inversion (i.e., $i < 0.6$), by carefully tuning synthesis temperature and cooling rate{\cite{Malavasi2002,Irani1962}} during MgMn$_2$O$_4$ synthesis.

Higher Mg conductivity, as is required for a solid state electrolyte, demands a lower cut-off for the migration barrier along a pathway. In the case of MgIn$_2$S$_4$, where we used a 500~meV cut-off, MC simulations indicate that the stoichiometric spinel should remain percolating up to $i \sim 0.44$. However, high degrees of inversion ($i \sim 0.85$) can be observed during MgIn$_2$S$_4$ synthesis (Figure~S1). As a result, strategies to limit inversion (i.e., $i < 0.44$) in sulfide spinel ionic conductors, such as chemical doping and careful calibration of synthesis conditions, need to be sought. 

\subsection{Voltages and capacities}
Inversion can also significantly impact electrochemical properties, such as phase behavior, average voltages and extractable capacities in an oxide-spinel cathode (Figure~{\ref{fig:6}}). For example, the average voltage for Mg intercalation, across x$_{\rm Mg} = 0 - 1$ in the Mn$_2$O$_4$-spinel, is higher in an inverted spinel compared to a normal spinel (Figure~{\ref{fig:6}}b). Mg intercalation experiments in spinel-Mn$_2$O$_4$ have reported a marginally higher average voltage ($\sim$~2.9~V){\cite{Kim2015a}} than predicted for the normal spinel ($\sim$~2.84~V), with extraction voltages as high as $\sim$~3.5~V during the charging cycle, which might be an indication of the spinel inverting during electrochemistry. Also, the calculated 0~K phase diagram of the Mg-Mn$_2$O$_4$ system (Figure~{\ref{fig:6}}a) suggests that the tendency to invert is the highest at an intermediate Mg concentration, as indicated by low $E_{formation}$ ($<$ 50~meV/Mn$_2$O$_4$) configurations with $i = 0.25$ at x$_{\rm Mg} = 0.5$. Hence, the degree of inversion in the Mn-spinel can indeed change dynamically during electrochemical Mg cycling, especially due to the presence of mobile Mn$^{2+}$ ions (Figure~S9). As reported by Ling \emph{et al.}{\cite{Ling2013}}, the mobility of Mn$^{2+}$ within the spinel can also depend on the local arrangement of Mg$^{2+}$ ions. Thus, from the data in Figure~{\ref{fig:6}}a, we expect the degree of inversion to vary largely between 0 and 0.25 during Mg-cycling. Also, previous Mg-cycling experiments in spinel-Mn$_2$O$_4$ have reported solvent co-intercalation based phase transformations,{\cite{Kim2015b,Sun2016}} which can be aided by the presence of mobile Mn$^{2+}$ ions. 

Additionally, the first Mg-site that will be (de-)intercalated in the spinel will depend on the degree of inversion on the surface of the cathode particle. For example, if the degree of inversion is $\sim$~0 on the surface, then Mg$^{2+}$ ions present in the $8a$ sites will be de-intercalated first from magnesiated-MgMn$_2$O$_4$. Similarly, in a partially inverted surface of a discharged cathode, the Mg$^{2+}$ ions in $16d$ sites that are connected via Hop~3 channels will be extracted as well as those in $8a$ sites connected via Hop~1 and open Hop~2 channels. In the case of a partially inverted surface in a charged-Mn$_2$O$_4$ cathode, the Mg$^{2+}$ ions are more likely to first insert into $16d$ channels connected via Hop~5, since Hop~5 exhibits lower Mg migration barriers compared to Hop~1 (Figure~{\ref{fig:2}}) at x$_{\rm Mg} \sim 0$.  

Since the percolation threshold in the oxide cathode can change with the vacancy concentration during Mg (de)intercalation (Figure~{\ref{fig:5}}a), a dynamic change in the degree of inversion during Mg-cycling can cause polarization within the cathode particle. For example, if $i$ changes from 0.55 to 0.59 while charging the MgMn$_2$O$_4$ cathode, during the following discharge the spinel is percolating only up to Mg$_{0.5}$Vac$_{0.5}$Mn$_2$O$_4$ (z~=~0.5 in Figure~{\ref{fig:5}}a) at $i = 0.59$. For further discharge into the structure, i.e., from Mg$_{0.5}$Vac$_{0.5}$Mn$_2$O$_4$ to  MgMn$_2$O$_4$, a reduction in $i$ to 0.55 is necessary, which can lead to hysteresis in the voltages during the charge and discharge cycles. Importantly, the extractable Mg content in stoichiometric MgMn$_2$O$_4$ decreases continuously with inversion, reaching values of $\sim$~63\% (171~mAh/g) and $\sim$~17\% (46~mAh/g) at $i = 0.25$ and 0.5 (Figure~{\ref{fig:6}}c), respectively. Thus, strategies to minimize changes in $i$, during Mg$^{2+}$ cycling, such as cation-doping of Mn to prevent Mn$^{2+}$ generation, should be employed to ensure reversible Mg (de)intercalation.


\section{Conclusion}
\label{sec:conclusion}
Spinels are promising materials in the development of multivalent battery electrodes and solid electrolytes but are prone to antisite disorder in the form of spinel inversion.
With the example of two prototypical oxide and sulfide spinels, MgMn$_2$O$_4$ (cathode) and MgIn$_2$S$_4$ (solid electrolyte), we demonstrated that inversion can significantly impact both Mg-ion mobility and electrochemical properties.
Using first-principles calculations, we analyzed the migration barrier for Mg$^{2+}$ hopping in different local cation arrangements, and found that inversion can both open and close select migration pathways on the atomic scale.
To quantify the influence of local barrier changes on the macroscopic transport of Mg$^{2+}$ ions, we determined the minimal M-deficiency x~in~Mg$_{\rm x}$M$_{\rm 3-x}$X$_4$ required for percolation. Using a cut-off of 750~meV and 500~meV for cathodes and solid electrolytes, respectively, we found that the stoichiometric MgMn$_2$O$_4$ and MgIn$_2$S$_4$ compositions are Mg percolating up to $\sim$~55--59\% and 44\% inversion.
Since the degree of inversion in the spinels considered in this work may vary between 20\% and 85\% depending on the method of preparation,\cite{Malavasi2002, Irani1962, Canepa2017a} a careful calibration of the synthesis conditions is essential to ensure sufficient Mg transport and to reduce the resultant impedance.
In addition, spinel inversion can affect the electrochemical properties of cathode materials by changing the phase behavior, average voltage, and extractable capacities.
Specifically, we find that the degree of inversion can change dynamically during electrochemical Mg cycling, as indicated by the 0~K phase diagram of the Mg$_{\rm x}$Mn$_2$O$_4$ system and the activation barriers for Mn$^{2+}$ hopping. Notably, even low degrees of inversion ($i < 0.4$) can detrimentally reduce the extractable capacity in stoichiometric MgMn$_2$O$_4$, with an estimated 15\% decrease in capacity with every 10\% increase in inversion. Thus, spinel inversion can hinder the electrochemical performance of both cathodes and solid electrolytes in MV systems and synthesis efforts must always be made to stabilize the normal spinel structure.  

Given that the Mg$^{2+}$ migration barriers over a range of oxide{\cite{Liu2015}} and sulfide spinels{\cite{LiuJainQuEtAl2016}} show similar trends, we expect similar behavior upon inversion in other spinel materials.
Finally, the framework developed in this work, particularly the data reported on percolation thresholds and extractable Mg, is readily transferable to other spinels that have potential applications in Li-ion, Na-ion, Ca/Zn-multivalent and other battery fields.


\begin{acknowledgements}
The current work is fully supported by the Joint Center for Energy Storage Research (JCESR), an Energy Innovation Hub funded by the U.S. Department of Energy, Office of Science and Basic Energy Sciences. This study was supported by Subcontract 3F-31144. The authors thank the National Energy Research Scientific Computing Center (NERSC) for providing computing resources. Use of the Advanced Photon Source at Argonne National Laboratory was supported by the U.S. Department of Energy, Office of Science, Office of Basic Energy Sciences, under Contract No.~DE-AC02-06CH11357. The authors declare no competing financial interests. GSG is thankful to Daniel C.\ Hannah at Lawrence Berkeley National Laboratory for a thorough reading of the manuscript.
\end{acknowledgements}

\bibliography{biblio}

\end{document}